\documentclass{emulateapj}
\shortauthors{Wang et al.}

\newcommand{\psrt}{PSR J1549$-$4848}
\newcommand{\psr}{J1549}
\newcommand{\spitzer}{\textit{Spitzer}}
\newcommand{\ha}{H$_{\alpha}$}

\begin{document}

\title{Serendipitous Discovery of An Infrared Bow Shock Near PSR
J1549-4848 with \textit{Spitzer}}

\author{Zhongxiang Wang\altaffilmark{1}, 
David L. Kaplan\altaffilmark{2},
Patrick Slane\altaffilmark{3},
Nidia Morrell\altaffilmark{4},
and Victoria M. Kaspi\altaffilmark{5} 
}

\altaffiltext{1}{\footnotesize 
Shanghai Astronomical Observatory, Chinese Academy of Sciences,
80 Nandan Road, Shanghai 200030, China}

\altaffiltext{2}{Physics Department, University of Wisconsin—Milwaukee, 
Milwaukee, WI 53211, USA}

\altaffiltext{3}{Harvard-Smithsonian Center for Astrophysics, 60 Garden Street, Cambridge, MA 02138, USA}

\altaffiltext{4}{Las Campanas Observatory, Observatories of the Carnegie 
Institution of Washington, La Serena, Chile}

\altaffiltext{5}{Department of Physics,
McGill University, 3600 University Street, Montreal, QC H3A 2T8, Canada}

\begin{abstract}
We report on the discovery of an infrared cometary nebula around
PSR J1549$-$4848 in our Spitzer survey of a few middle-aged radio pulsars.
Following the discovery, multi-wavelength imaging and spectroscopic 
observations of the nebula were carried out. We detected the nebula 
in Spitzer IRAC 8.0, MIPS 24 and 70 $\mu$m imaging and in Spitzer 
IRS 7.5--14.4 $\mu$m spectroscopic observations, and 
also in the WISE all-sky survey at 12 and 22 $\mu$m.
These data were analyzed in detail, and we find that the nebula can be 
described with a standard bow-shock shape, and that its spectrum contains
polycyclic aromatic hydrocarbon and H$_2$ emission features. However, 
it is not certain which object 
drives the nebula. We analyze the field stars and conclude that none of them 
can be the associated object because stars with a strong wind or mass 
ejection that
usually produce bow shocks are much brighter than the field stars. 
The pulsar is approximately 15\arcsec\ away from the region in which the 
associated object is expected to be located. In order to resolve 
the discrepancy,  
we suggest that a highly collimated wind could be emitted from the pulsar and
produce the bow shock. 
X-ray imaging to detect the interaction of the wind with the ambient medium
and high-spatial resolution radio imaging to determine the proper motion 
of the pulsar should be carried out, which will help verify the association of 
the pulsar with the bow shock nebula.  

\end{abstract}


\section{INTRODUCTION}

Interstellar shocks are seen in association with various 
astrophysical objects, from wind-blowing massive stars \citep{van+90}, 
mass-ejecting giant stars (e.g., \citealt{mar+07}),
expanding supernova remnants (SNRs; e.g., \citealt{mor+07}), to 
energetic pulsars \citep{gs06}. In these objects, fast
moving ejecta collide with the surrounding interstellar medium (ISM), 
and the strong interaction drives a shock front. Depending on the kinetic 
energy carried into the shock and local ISM conditions, the postshock hot 
material can be bright at different wavelengths. Observations of shocks
not only help study shock physics, but also provide probes of the associated 
objects and ISM.

Depending on temperature $T$, the sound speed $c_s$ in the ISM 
[$c_s\simeq 10(T/10^4\ {\rm K})^{1/2}$ km s$^{-1}$] is in the range of 
1--100 km s$^{-1}$. As a result, when stellar objects have high space 
velocities $V_{\ast}$, they can be moving supersonically in the ISM. 
In such cases, a shock will appear as a cometary structure 
(so-called bow shock), as interstellar material is swept up into a dense 
cone-type shell (e.g., \citealt{wil96}). 

One type of bow shock is known to be associated with pulsars. Having space 
velocities of $\geq$100 km s$^{-1}$ (e.g., \citealt{hob+05}) and strong winds 
(which carry much of the spin-down energy $\dot{E}$ of pulsars), 
pulsars moving in the ISM are candidates to drive bow shocks.
Thus far, six pulsars have been found to have a typical bow shock,
revealed by the shocks' optical Balmer line emission \citep{gs06}. 
The detected line nebulae are understood to arise from 
de-excitations of neutral H atoms in the shocked ambient gas, 
following the processes of collisional excitation or charge-exchange 
(e.g., \citealt{bb01}). In addition, a pulsar bow shock
may also be revealed by the detection of a termination shock of 
the pulsar wind. The termination shock lies inside a bow shock, and can appear
bright at X-ray or radio energies due to synchrotron radiation 
(\citealt{krh06}; \citealt{gs06}).

In our \spitzer\ Infrared Array Camera (IRAC) survey of 7 relatively 
young pulsars, for the purpose of probing the general existence of 
debris disks around isolated neutron stars (Wang et al. 2013, in preparation), 
a mid-infrared (MIR) cometary nebula was serendipitously 
detected near one of the pulsars, \psrt\ (hereafter \psr). Naively thinking 
that the MIR 
structure is associated with the pulsar, we would have found the first MIR bow 
shock that arises from the interaction between the pulsar wind and ISM. 
We thus made follow-up 24 and 70~$\mu$m imaging and 7.5--14.4~$\mu$m 
spectroscopic observations with \spitzer\ and optical observations 
with the Magellan telescopes after the initial detection of the nebula 
at 8.0 $\mu$m. However, our detailed data analyses raise questions 
about the association between the cometary nebula and the pulsar.
On the basis of our observational results, the origin of the nebula  
is not certain. Here we report our multi-wavelength observations, 
data analyses, and results. 
The observations and data reduction are described in \S~\ref{sec:obs}, 
and the data analyses and results are presented in \S~\ref{sec:ana} 
and \S~\ref{sec:res}, respectively.
We discuss the possible origin of the cometary nebula in \S~\ref{sec:disc}
and summarize the results in \S~\ref{sec:sum}.
\begin{figure*}
\begin{center}
\includegraphics[scale=0.95]{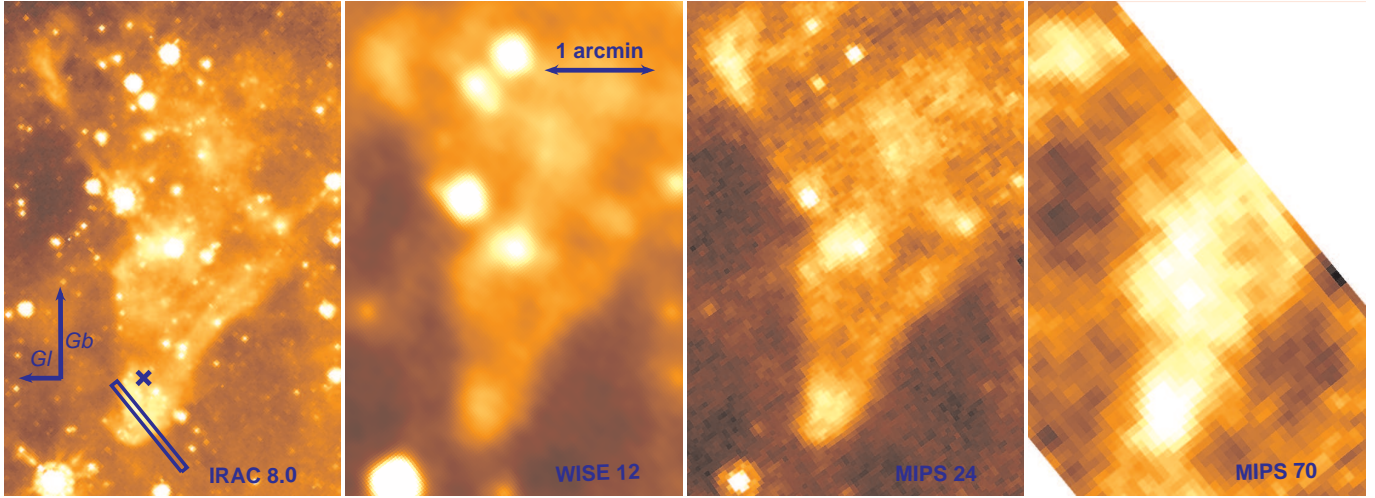}
\figcaption{\spitzer\ IRAC 8.0 $\mu$m (left panel) and 
MIPS 24 (third panel) and 70 $\mu$m (right panel) imaging 
of the cometary nebula. The WISE third-band 12 $\mu$m detection of the nebula
is shown in the second panel. In the left panel,
the position of \psr\ is marked by a cross sign, and one nod position
of the slit (which has a size of 57\arcsec$\times$3\farcs7)  
in our \spitzer\ IRS observations is indicated with a box region.
\label{fig:field} }
\end{center}
\end{figure*}

\section{OBSERVATIONS AND DATA REDUCTION}    
\label{sec:obs}

Our \spitzer\  and ground-based observations of the cometary nebula are
summarized in Table~\ref{tab:obs}. Below we describe them and the
related data reductions.

\subsection{\spitzer\ Imaging}

\subsubsection{IRAC And MIPS Imaging}

As part of our survey program, the initial \spitzer\ observations of \psr\ were
carried out on 2007 September 12. The imaging instrument used was 
the Infrared Array Camera (IRAC; \citealt{fha+04}). 
The field of our target was imaged in the pair of 4.5 and 8.0 $\mu$m 
channels of the IRAC simultaneously.
The detectors at the short and long wavelengths were InSb
and Si:As devices, respectively, with 256$\times$256 pixels and a plate
scale of 1\farcs2/pixel. The field of view (FOV) was 5\farcm2$\times$5\farcm2.
The frame time was 30 s, with 26.8 effective exposure time
per frame. The total exposure time at each channel was 17.87 min.
A cometary nebula was detected in the 8.0 $\mu$m image 
(Figure~\ref{fig:field}) around \psr, but not in the 4.5 $\mu$m image.

To study the nebula and derive its properties, we requested
\spitzer\ observations at the longer wavelengths. 
Using Multiband Imaging Photometer for \spitzer (MIPS; \citealt{rieke+04}),
24 and 70 $\mu$m broad-band imaging of the source field was carried out 
on 2008 August 30.
The 24 $\mu$m detector was a 128$\times$128 pixel
Si:As array, with a pixel size of 2\farcs55 and a 5\farcm4$\times$5\farcm4
FOV. The 70 $\mu$m detector was a 32$\times$32 pixel Ge:Ga array.
We chose its wide FOV observing mode for imaging, which provided a pixel scale
of 9\farcs85$\times$10\farcs06 and a 5\farcm2$\times$2\farcm6 FOV.
At 24 and 70 $\mu$m, the exposure times were 11.8 and 10.5 min, respectively
(frame times $\simeq$ 3 sec). 

\subsubsection{Data Reduction}

We started with the Basic Calibrated Data (BCD), provided by \spitzer\ Science
Center (SSC). The data were produced from raw images through the IRAC and
MIPS data pipelines (IRAC software version was S16.1.10 and MIPS software
version was S18.1.0) in SSC. The detailed reduction in the pipelines can be
found in the IRAC Data Handbook (version 3.0) and MIPS data Handbook 
(version 3.3.1). 

For IRAC data, we first used the IRAC artifact mitigation program, distributed
as Contributed Software by SSC, to clean the BCD images. Several 
bright stars in the target field caused column pull-down, row pull-up, 
or optical banding artifacts. 
We then corrected each BCD image for array location dependence, using the 
array location correction images provided by SSC. To combine BCD images
into a final post-BCD (PBCD) mosaic at each channel, we used 
{\tt MOPEX}, an SSC's package for reducing and analyzing imaging data.

For MIPS 24 $\mu$m data, the BCD images do not have any artifacts that
need corrections. We used {\tt MOPEX} to combine the BCD images
into a final PBCD mosaic. The MIPS 70 $\mu$m BCD images contain horizontal
and vertical strips, which were due to response variations and stim latents 
(stims were
MIPS's built-in calibration light sources, which flashed during an observation
sequence; for details, see MIPS Data Handbook), respectively. 
We used the GeRT package, which is provided for MIPS 70 \& 160 $\mu$m
data reductions, to clean up the BCD images. During the reduction, the region
that contains our source was masked out. The BCD images were combined
into one final PBCD mosaic using {\tt MOPEX}.

\subsection{\spitzer\ Spectroscopy}

\subsubsection{IRS Observations}

Our \spitzer\ spectroscopic observations were made on 2009 September 14.
The instrument was the Infrared Spectrograph (IRS; \citealt{hou+04}).
We used the Short-Low (SL) module of the IRS for the observations,
which provided a wavelength coverage of
7.5--14.4 $\mu$m (first order) and a resolving power of 61--120.
The detector was a 128$\times$128 arsenic doped silicon (Si:As) array,
with a pixel scale of 1.8\arcsec/pixel. The slit in the first order had a 
length$\times$width of 57\arcsec$\times$3\farcs7.
We obtained a set of 16 spectra at each of two default nod positions along
the slit. The integration time for each 
individual spectrum was 60.95 s, and the total on-source time was 32.5 min.
In addition, since our target is an extended source,
we used the exact same observing mode and obtained spectra of a nearby 
sky region for background subtraction. 

\subsubsection{Data Reduction}

We started with the BCD images, produced from the SSC IRSX Channel-0 Software
(version S18.7.0). We first combined all the sky BCD images 
into one master sky image and subtracted this master image from each of
the target BCD images. We then used the {\tt IRSCLEAN\_MASK} program,
provided by SSC, to construct a super mask by combining the rogue 
pixel mask for the campaign IRSX009500 (in which our IRS observations were 
made) and additional bad pixels identified by {\tt IRSCLEAN\_MASK}. 
The super mask 
was used in the {\tt IRSCLEAN} program for cleaning up bad pixels 
in the BCD images.
The BCD images were combined into one final image by using the 
{\tt coa2d.pro} program.

We used {\tt SPICE}, an IRS spectrum extraction software provided by SSC,
to extract the spectrum of the nebula. The width of the extraction box
was 15 pixels (27\arcsec), which covers the 2-D spectrum image well.
The spectra at two nod positions were extracted and average-combined
into the final source spectrum.

\subsection{WISE Imaging}

Launched on 2009 December 14, the Wide-field Infrared Survey Explorer (WISE) 
mapped the entire sky at 3.4, 4.6, 12, and 22 $\mu$m (called W1, 
W2, W3, and W4 bands, respectively) in 2010 with FWHMs of 
6.1\arcsec, 6.4\arcsec, 6.5\arcsec, and 12.0\arcsec\  in the four bands, 
respectively (see \citealt{wri+10} for details). 
The WISE all-sky images and source catalogue were released in 2012 March.
We downloaded the WISE image data of the source field 
from the Infrared Processing and Analysis Center (IPAC),
and found that the cometary nebula was detected in the W3 and W4 bands but 
not in the first two bands.  The WISE observations were made between 
2010 Feb. 22-25 and the depth of coverage 
was $\sim$13.7 pixels (corresponding to 110~s on-source integration time).

\subsection{Ground-Based Observations}

\subsubsection{Optical/Near-Infrared Imaging}

We also observed the target field in $R$ broad band and H$_{\alpha}$
narrow band on 2008 June 15, using Inamori-Magellan Areal Camera and 
Spectrograph (IMACS) on the 6.5-m Baade Magellan Telescope at Las Campanas 
Observatory in Chile. The long camera (known as f/4) of IMACS was used. 
The detector of the camera consists of 8 2048$\times$4096 pixel$^2$ SITe
ST-002A CCDs. Under the f/4 imaging mode, the camera provides a field coverage
of 15\farcm4$\times$15\farcm4 and a pixel scale of 0\farcs111 pixel$^{-1}$.
The exposure times of the images were 4 min in $R$ and
18.3 min in \ha. The observing conditions were good, with seeing 
being 0\farcs65.
In addition, we observed standard stars PG1633+099 \citep{lan92} 
in $R$ band and LTT7379 \citep{ham+94} in \ha\ for flux calibration. 

We included near-infrared (NIR) $K_s$ imaging data of the target field, obtained
on 2006 May 17 as part of our IR survey of a few young pulsars (like \psr).
The observation was also made with the Baade Magellan Telescope.
The NIR camera was Persson's Auxiliary Nasmyth Infrared Camera 
(PANIC; \citealt{mpm+04}).The detector was a Rockwell Hawaii 
1024$\times$1024 HgCdTe array, having a field of view (FOV) 
of 2\arcmin$\times$2\arcmin\  and a pixel scale of 
0\farcs125 pixel$^{-1}$. The total on-source exposure time was 
22.5 min. During the exposure, the telescope was dithered in a
3$\times$3 grid with offsets of 10\arcsec\ to obtain a measurement of
the sky background. The observing conditions were good, with 0\farcs4 
seeing in $K_s$.

\subsubsection{Data Reduction}

We used the IRAF data analysis package for data reduction.
The images were bias-subtracted and flat-fielded. In addition for
the NIR data, a sky image was made by filtering out stars
from each set of dithered images in one observation.
The sky image was
subtracted from the set of images, and then the sky-subtracted images
were shifted and combined into one final image of the target field.

\section{ANALYSIS}
\label{sec:ana}

\subsection{Astrometry}

The pointing of the IRAC frames is typically accurate to 0\farcs5.
In order to determine more accurately the position of the cometary nebula and
locate \psr\ on the images, we astrometrically calibrated the IRAC
8.0 $\mu$m image by matching stars detected in this image 
to Two Micron All-Sky Survey (2MASS; \citealt{2mass}) stars.
For this calibration, 190 2MASS stars were used.
The nominal uncertainty of the calibrated image 
is dominated by the 2MASS systematic uncertainty ($\simeq$0\farcs15, with
respect to the International Celestial Reference System).
The other images were also positionally calibrated by matching them 
to the 8.0 $\mu$m image.

\subsection{Morphology}
\label{subsec:mor}

\begin{figure*}
\begin{center}
\includegraphics[scale=0.95]{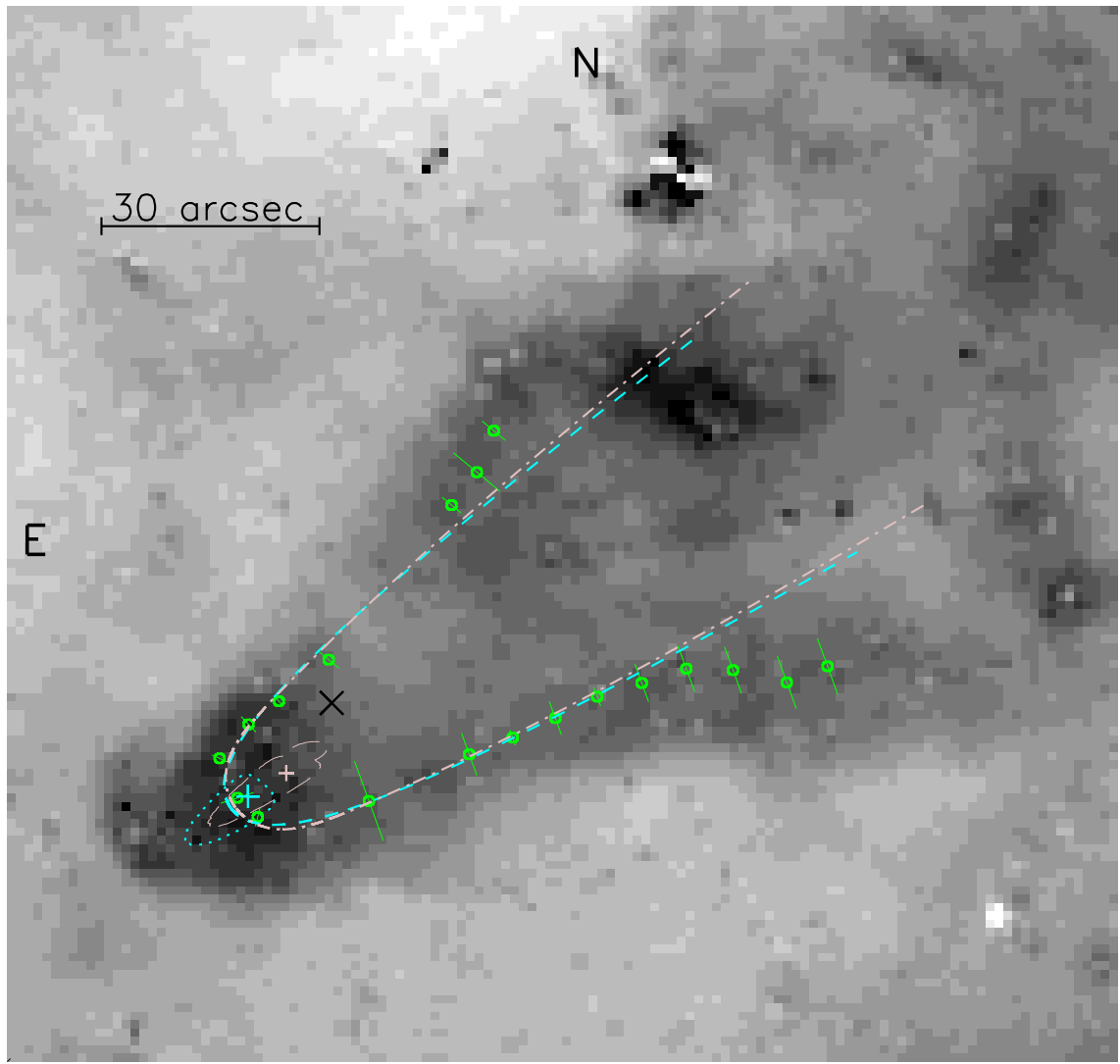}
\figcaption{\spitzer\ IRAC 8.0 $\mu$m PBCD image of the cometary nebula with
in-field stars removed. The small circles mark the brightness peak points
we obtained for the nebula (see Section~\ref{subsec:mor}). 
The position of the pulsar \psr\ is marked by a black cross sign. 
The dotted curve indicates the 3$\sigma$ region 
in which the associated object is considered to be located 
(obtained from fitting the analytic solution for bow shocks provided 
by \citealt{wil96}), with the large plus sign marking the best-fit position
and dashed curve being the best-fit.
The long dashed curve indicate the same region obtained from fitting 
the analytic solution for pulsar bow shocks provided by \citet{vig+07}, with
the small plus sign marking the best-fit position and dot-dashed curve
being the best-fit.
\label{fig:bow} }
\end{center}
\end{figure*}

We studied the cometary structure by semi-quantitatively determining its shape.
The 8.0 $\mu$m image was used for the detailed study because it has
the best resolution among the \spitzer\ and WISE detections. We first removed 
all point
sources near the cometary structure. Using the software package {\tt APEX} 
(multi-frame extraction) provided by SSC, point sources in the field were 
identified and subtracted from each BCD image. The residual frames were 
then combined into a PBCD mosaic, which is shown in Figure~\ref{fig:bow}.

We then manually selected small regions along the three parts of the nebula:
the two wings and the head region. Narrow box regions set parallel to 
the three nebula parts were defined, with each box having a length$\times$width 
of 5$\times$1 pixel$^2$ (6\farcs1$\times$1\farcs2). 
Photons within each box were counted and the uncertainty
on the photon counts was estimated as the square-root of the total counts.
This way a brightness curve that was perpendicular to the narrow box regions
(or to the nebula parts) was obtained.  Using a Gaussian function, 
we performed least-squares fitting to the brightness curve and thus determined
the location of a brightness peak point. The uncertainty for each location 
was also estimated from the fitting. In total 19 such brightness peak points
were determined (see Figure~\ref{fig:bow}).

\subsection{Photometry}

We obtained the flux of the head region of the cometary nebula at each band.
A circular region with a radius of 21\arcsec, 
which well covers the head region, was used as the photometry aperture. 
A box region of 80\arcsec$\times$50\arcsec\ near the head was used 
for estimating the sky background brightness. 
For photometry at 8.0 $\mu$m,
aperture correction was required because IRAC imaging uses
point sources for flux calibration. We used the function form provided
by SSC
to calculate the correction factor and found a value of 0.83 for a radius of
21\arcsec. The flux value at 8.0 $\mu$m is the measured flux value times the
correction factor. We also derived the flux upper limit
for the 4.5 $\mu$m non-detection, because the deep image provides an 
additional constraint
on MIR emission from the cometary nebula. 

Following the guidelines given for aperture photometry on the WISE atlas 
images\footnote{http://wise2.ipac.caltech.edu/docs/release/allsky/expsup/sec2\_3f.html}, 
WISE W3 and W4 fluxes were also obtained. The broad-band flux
measurements and 4.5 $\mu$m upper limit are given in Table~1.
\begin{figure*}
\begin{center}
\includegraphics[scale=0.85]{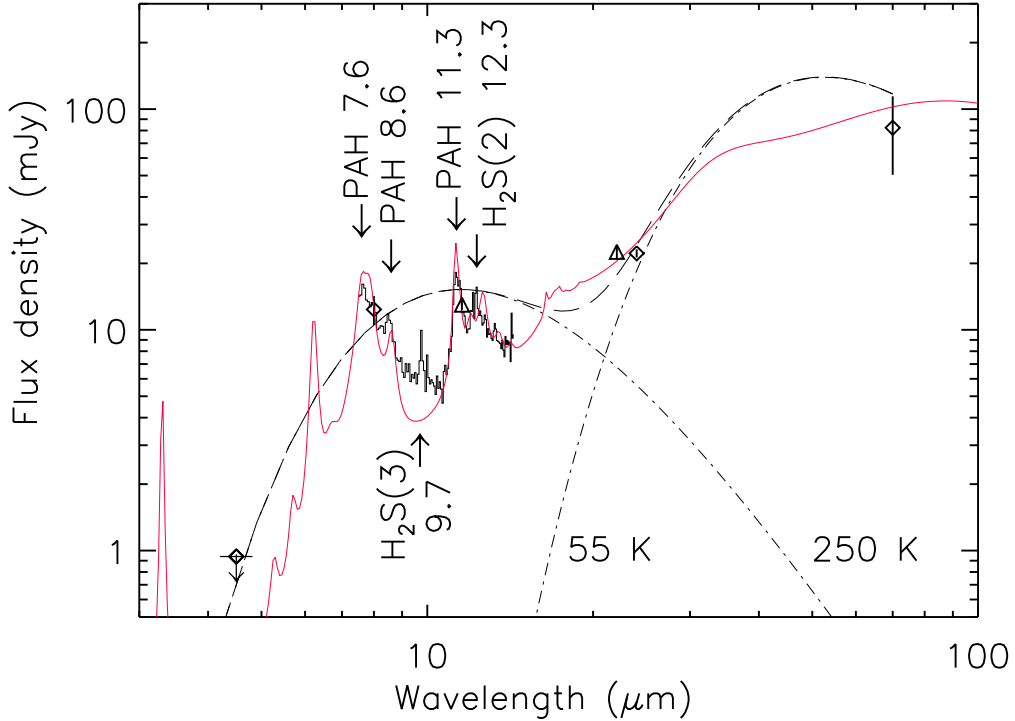}
\figcaption{Dereddened spectrum of the head region of the cometary nebula,
which consists of the \spitzer\ broad-band 4.5 $\mu$m flux upper limit, 
8.0, 24, and 70 $\mu$m fluxes (diamonds) and IRS 7.5--14.4 $\mu$m spectrum 
(solid histogram), and WISE 12 and 22 $\mu$m fluxes (triangles). 
Dust emission from two temperature components (250~K and 55~K, both
dot-dashed curves; their sum, long-dashed curve) can generally describe
the spectrum. The model spectrum of irradiated dust emission given 
by \citet{dl07} is plotted as the red solid curve.
\label{fig:spec} }
\end{center}
\end{figure*}
 
\section{RESULT}
\label{sec:res}

\subsection{Location of the Associated Object}
\label{subsec:loo}

The shape of the IR cometary nebula indicates that it 
is likely a typical bow shock. The location of the associated object that 
drives the IR nebula can thus be estimated by fitting the
nebula with a standard 
bow-shock shape, since the distance from an associated object to the apex of the
associated bow shock (so-called ``standoff distance") sets the length 
scale of the bow shock. We considered an analytic solution derived 
by \citet{wil96}, 
$r_{\theta}\sin\theta = r_a\sqrt{3(1-\theta/\tan\theta)}$, where $r_{\theta}$
is the distance between the bow shock shell and the associated 
object, $\theta$ is
the polar angle from the axis of symmetry (i.e., the associated object's 
moving direction), and $r_a$ is the standoff distance. With this solution,
we fit the positions of the brightness peak points obtained 
in \S~\ref{subsec:mor}. 
The free parameters in the fitting were $x$, $y$ location of the associated
object,
$r_a$, and the rotation angle of the axis of the bow shock in the plane
of the sky. To simplify the fitting, we assumed that 
the axis of symmetry of the bow shock lies nearly in the plane 
of the sky (note that an inclination angle such as 60\arcdeg\ will
result in a smaller projected distance between the associated object 
and the `apparent' apex of the bow shock).

The obtained (3$\sigma$) region in which the associated object is located
is indicated as a dotted contour 
in Figure~\ref{fig:bow}. It is a small region very close to
the head of the cometary nebula, and approximately 15\arcsec\ away
from \psr\ (whose position is R.A. = 15$^{\rm h}$49$^{\rm m}$21\fs15, 
Decl. = $-$48\arcdeg48\arcmin37\farcs4, equinox J2000.0, and has an 
uncertainty of $\simeq$1\arcsec; \citealt{dam+98}).  
Because the data points of the brightness
peaks do not provide tight constraints along the axis of symmetry, 
the region is more extended along this direction. The best-fit location and 
bow shock solution
are shown as a plus sign and a dashed curve, respectively, in
Figure~\ref{fig:bow}. As can be seen,
the standard bow shock shape generally fits the data points, 
although the tail of the nebula is wider, 
deviating away from the model. 
From the best fit ($\chi^2$=14 for 15 degrees of freedom), 
we found $r_a\simeq 2\farcs9$. 

\subsection{Mid-IR Spectrum}

In Figure~\ref{fig:spec}, we show the broad-band and \spitzer\ IRS spectrum 
of the head region of the cometary nebula, which is dereddened with
$E(B-V)=1.1$ \citep{sfd98} using the interstellar extinction laws given
by \citet{ind+05} for wavelengths $\leq$8 $\mu$m and by \citet{wd01} for
wavelengths $>$8 $\mu$m. The \spitzer/IRAC 4.5 $\mu$m flux upper limit is
also included. The IRS spectrum was flux calibrated by equaling
the average value of its 7.5--8.5 $\mu$m region
to the IRAC 8.0 $\mu$m broad-band flux value. 
A few emission lines were detected in the IRS spectrum, which are 
polycyclic aromatic hydrocarbon
(PAH) 11.3, 8.6, and likely 7.6 $\mu$m. These PAH lines are commonly seen 
in many emission nebulae and arise from vibrational modes of PAH in dust 
(e.g., \citealt{dra03}).
Two H$_2$ rotational lines were also detected, although weakly. 
They are H$_2$S(3) 9.7 $\mu$m and H$_2$S(2) 12.3~$\mu$m. In general, H$_2$ 
lines are associated with photodissociation regions (PDRs; e.g., \citealt{ht97})
or shock interaction regions (e.g., \citealt{shi+09}).
We note that the broad dip at $\sim$9.7 $\mu$m is not the well-known
silicate absorption feature. Rather, its appearance is caused by the nearby
strong  PAH emission features \citep{dl07}. 
The total Galactic reddening along the line of sight to the source
is only approximately $E(B-V)=1.1$ (for $l$=330\fdg5, $b$=4\fdg3; 
\citealt{sfd98}). If the dip is caused by 
interstellar reddening, $E(B-V)$ should be $\geq 4$, which is too large to
be consistent with those estimated from the all-sky dust \citep{sfd98}
or neutral hydrogen density \citep{dl90} maps. 
Also, as shown below in this section,
the local column density of the cometary nebula is much lower than that 
implied by the reddening $E(B-V)\geq 4$.
\begin{figure*}
\begin{center}
\includegraphics[scale=0.85]{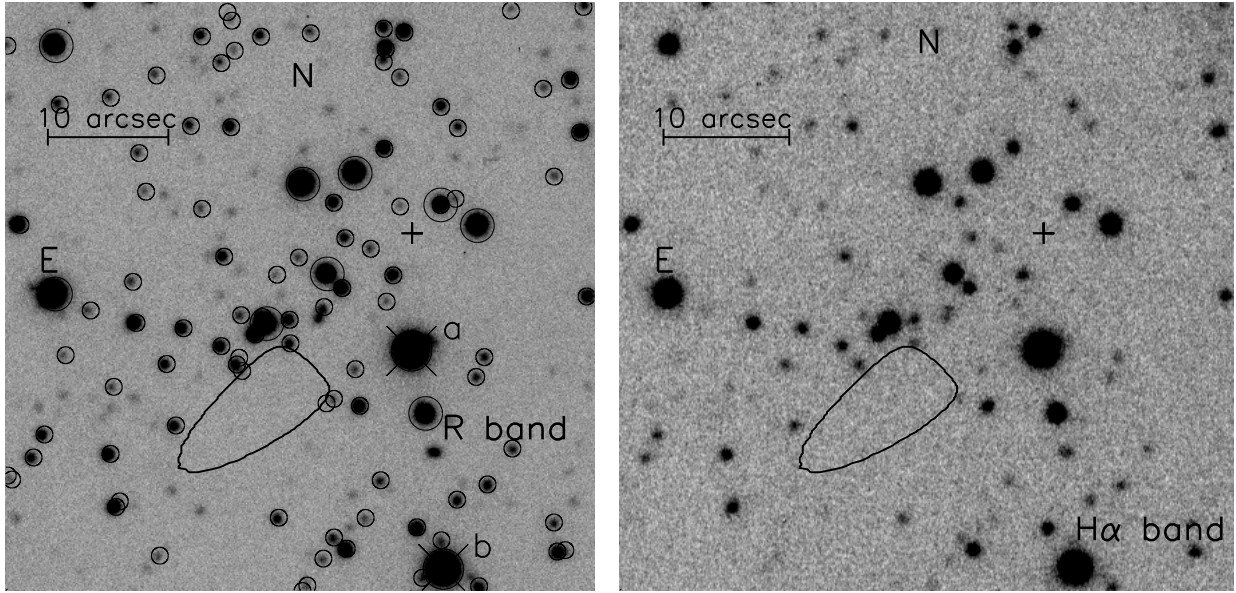}
\figcaption{Optical $R$ ({\it left} panel) and $H_{\alpha}$ ({\it right} panel)
images of the field of the IR cometary nebula. In each panel, 
the region in which 
the associated source should be located is indicated by the solid curve, 
while the position of \psr\ is indicated by the plus sign.  
Photometry of the field stars (marked by circles in the left panel) was 
performed, and
star $a$ and $b$ (marked by cross signs) was slightly saturated in the $R$
and $K_s$ image, respectively.
\label{fig:bowrh} }
\end{center}
\end{figure*}

We first tested using a simple dust grain model \citep{dw81} to 
reproduce the spectrum we obtained. This model was developed to explain 
the observed IR emission from supernova remnants (SNRs), in which dust grains 
are heated both by collisions with ambient thermal gas and by radiation flux,
and could work for the IR nebula in our case. Considering the general case
that the grains' absorption/emission efficiency decreases as $\lambda^{-2}$,
we found that two components with temperatures of
$T_h=250$ K and $T_l=55$ K (assuming a simplified, uniform grain size of 
radius 0.1 $\mu$m),
can generally describe the spectrum (Figure~\ref{fig:spec}).
The dust masses thus simply estimated are
1.0$\times 10^{25}d^{2}_{\rm kpc}$~g and 
1.9$\times 10^{29}d^{2}_{\rm kpc}$~g, respectively, where $d_{\rm kpc}$
is the source distance in units of kpc. 

To better reproduce the IRS spectrum, a detailed model including 
PAH particles is needed.
We considered the model spectra derived and provided by \citet{dl07}, 
although we note that in their model, dust grains are heated by a field of 
starlight.
According to their calculations, for the case of dust grains
being heated by a single 
radiation intensity, the flux ratio between 24 and 70 $\mu$m 
is mainly sensitive to radiation with energy density per unit
frequency $UU_{\rm ISRF}$, 
where $U_{\rm ISRF}$ is the interstellar radiation field (ISRF) estimated
by \citet{mmp83} for the solar neighborhood and $U$ is a dimensionless 
scaling factor. We thus found that $U=0.1$ is required for our data points. 
However,
the $U=0.1$ model spectra have much stronger PAH features than those in
our IRS spectrum (even with the lowest PAH mass fraction $q_{\rm PAH}=0.47\%$;
$q_{\rm PAH}\simeq 4.6\%$ is found to be applicable to the dust 
in the Milky Way). In order to have
a good fit to the IRS spectrum, we found that the dust should be
heated by a single radiation intensity, $U=12$, plus by a power-law 
distribution of starlight intensities ranging from $U=12$--10$^6$ 
(where $q_{\rm PAH}\simeq 4.6\%$, the power-law index $\alpha$ is 
fixed, $\alpha=-2$, and the fraction of the dust mass that is
exposed to the distribution of starlight intensities is $\gamma = 0.16$;
cf. equation (23) in \citealt{dl07}). This model spectrum
is displayed in Figure~\ref{fig:spec}, which generally describes
the PAH features in our spectrum. 
As a result, the dust mass $M_{\rm dust}$ can be estimated from 
the total value of the 24, 70, and 160 $\mu$m fluxes (cf. equation (34)
in \citealt{dl07}). Since 
no flux measurement at 160 $\mu$m was obtained, 
$M_{\rm dust}$ should be $\geq 1.2\times 10^{29} d_{\rm kpc}^2$~g 
and $\simeq 1.9\times 10^{29} d_{\rm kpc}^2$~g when assuming
the flux at 160 $\mu$m is equal to that at 70$\mu$m. The value range is
consistent with that derived above from the simple dust grain model.

We further estimated the properties of the head region of the bow shock.
Considering again the dust size of 0.1 $\mu$m (and density of 3 g~cm$^{-3}$;
\citealt{dw81}), 
the total luminosities in the region were
5.7$\times 10^{31}d^2_{\rm kpc}$ and 
1.1$\times 10^{32}d^2_{\rm kpc}$ erg~s$^{-1}$ 
for the 250~K and 55~K dust grains, respectively. 
Simply assuming that the head region is a 21\arcsec\ radius 
(or 0.1$d_{\rm kpc}$ pc) sphere, the number density (dominated by the
cold dust grains) can be estimated to be $\sim$90 cm$^{-3}$, where
a mass ratio of 1 to 100 between the dust to gas is used \citep{dra+07}.

\subsection{H$\alpha$ Non-Detection}
\label{subsec:ha}

In order to search for possible line emission from the bow shock,
we made both $R$ and $H_{\alpha}$ imaging observations 
of the target field with the Magellan telescopes; the wide-band image 
was used to serve as the continuum flux.
We did not detect any $H_{\alpha}$ nebula in our observations. 
Both optical $R$- and $H_{\alpha}$-band images of the target field are shown
in Figure~\ref{fig:bowrh}. The $R$-band detection limit was $\sim$23 mag, and 
the $H_{\alpha}$ flux upper limit (3$\sigma$) was 
2.9$\times 10^{-14}$ erg~s$^{-1}$~cm$^{-2}$arcsec$^{-2}$.

\section{Discussion}
\label{sec:disc}
Provided with the results we have obtained from the above analyses, we discuss
the possible origin of this bow shock by first considering a stellar object
and then \psr\ as the associated star.
\begin{center}
\includegraphics[scale=0.65]{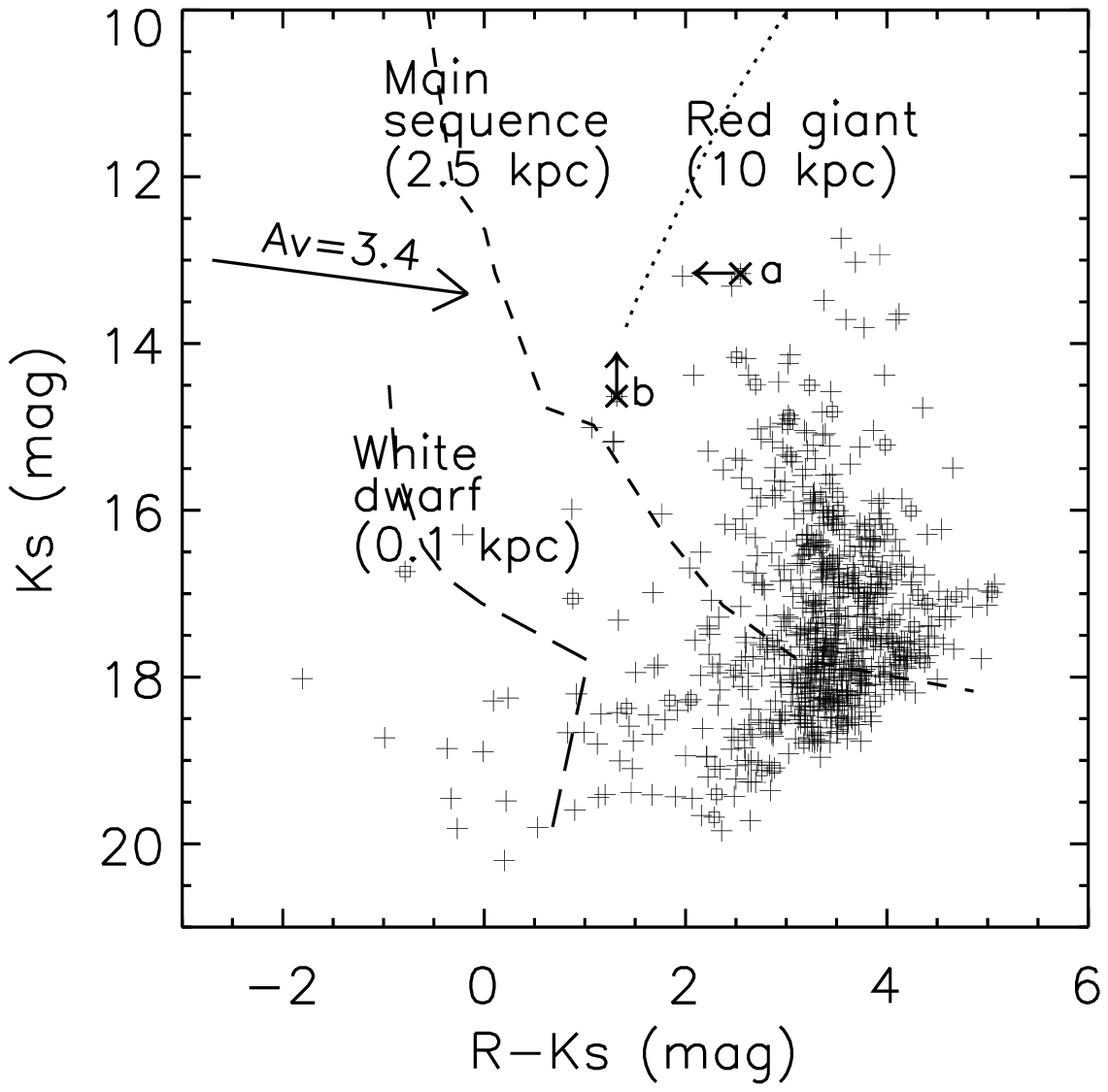}
\figcaption{$K_s$ versus $R-K_s$ diagram of the in-field stars
near the bow shock. Data points marked by squares are those stars
highlighted by circles in the left panel of Figure~\ref{fig:bowrh}.
The total Galactic extinction of $A_V=3.4$ is indicated by the arrow.
The two stars marked with arrows are those two nearby stars slightly 
saturated in our $R$ or $K_s$ images (Figure~\ref{fig:bowrh}), 
with the arrows indicating their magnitude or color limit.
Color-magnitude curves for main-sequence stars, red giants, and
white dwarfs are plotted as the dashed, dotted, and long-dashed lines.
\label{fig:rk} }
\end{center}

\subsection{A Stellar Bow Shock?}
As shown by our analysis in \S~\ref{subsec:loo}, considering an isotropic
wind/outflow from an object \citep{wil96}, the location of the associated object
would likely be within a small area near the head region of the bow 
shock, which is more than 15\arcsec\ away from PSR~\psr.
Such IR bow shocks 
can be driven by either massive stars with a strong 
wind \citep{van+90,pbb+12} or giant stars ejecting 
mass \citep{mar+07,cox+12},
as long as the associated stars are moving supersonically in the ISM. 
The stand-off distance of such a bow shock is given by
$r_a=(\dot{m}_w V_w/4\pi\rho V^2_{\ast})^{1/2}$ (e.g., \citealt{wil96}), where
$\dot{m}_w$ is the mass loss rate, $V_w$ is the wind velocity, and 
$\rho$ is the density of the ISM. 
For a massive O/B star with
typical values of $\dot{m}_w=10^{-7}\ M_{\sun}$~yr$^{-1}$, 
$V_w=1000$~km~s$^{-1}$,
and $V_{\ast}=30$~km~s$^{-1}$ (see, e.g., \citealt{pbb+12}),
the required ambient ISM hydrogen density to produce the observed bow shock is
$n_{H} \simeq 1500(\dot{m}_{w-7}V_{w3}/V^{2}_{\ast 30} d^{2}_{\rm kpc})$ cm$^{-3}$, 
where $\dot{m}_{w-7}$ is the mass loss rate in units 
of $10^{-7}\ M_{\sun}$~yr$^{-1}$, $V_{w3}$ is the wind velocity in units of
1000~km~s$^{-1}$, and $V_{\ast 30}$ is the star's space velocity in units
of 30~km~s$^{-1}$.
Here we have used $r_a=2\farcs9 d= 4\times 10^{16} d_{\rm kpc}$~cm and 
$\rho=\mu m_{H}n_{H}$ with $\mu_{H}=1.4$ ($\mu_{H}$ is the mean nucleus 
number per hydrogen atom and $m_{H}$ is the mass of the hydrogen
nucleus). Using the low end value of
$\dot{m}_{w-7}=0.1$ and $d_{\rm kpc}=10$ (see below for a discussion
of the distance), along with the other mass-loss parameters given above,
we find $n_{H}\approx 1$, broadly consistent with values expected for the
ISM density. Alternatively, for a giant star with mass
ejection, $V_w\simeq 10$~km~s$^{-1}$ and other parameters similar to those
above (e.g., \citealt{cox+12}), the required ambient density is 
$n_{H} \simeq 15$; reasonable adjustments to the parameters
can easily lead to values more consistent with the ISM density. 
These calculations thus show that it is possible to have either a 
massive star or
a giant star with mass ejection as the object that drives the bow shock.

However we evaluated the possible association of the bow shock
with any in-field stars and obtained negative results. 
First there were no stars detected in the associated
object region in our optical images down to $R\simeq  23$
(see Figure~\ref{fig:bowrh}). We also analyzed the nearby stars by 
constructing a color-magnitude diagram of $K_s$ versus $R-K_s$ 
(Figure~\ref{fig:rk}). As can be seen, the stars are consistent with
being low-mass main sequence stars (or white dwarfs) at distances of 
$\lesssim$2.5 kpc. Massive or giant stars 
would have to be at distances of $\sim$6 kpc and 
$\sim$10 kpc, respectively, and approximately 0.5--0.8 kpc away
from the Galactic plane, a place not likely to form bow shocks due to 
the low density of the ISM. The detection of H$_2$ lines
in the IRS spectrum suggests that the associated object should probably 
be $\lesssim 0.1$ kpc away from the Galatic plane, since molecular gas 
generally inhabits within the mid-plane of the Galactic disk \citep{cox05}.

\subsection{PSR \psr\ as the Associated Object?}
Given their high space velocities ($\geq$100 km s$^{-1}$) and strong winds,
pulsars can naturally drive bow shocks in the ISM. Detections of
bow shocks around pulsars at multiple wavelength ranges have provided
sufficient evidence for it \citep{krh06,gs06}. The existence of
IR bow shocks, arising due to irradiation or collisional heating,
can be expected.  The pulsar \psr\ has a spin-down energy 
of $\dot{E}=2.3\times 10^{34}$ erg s$^{-1}$ and a distance of 1.5 kpc
\citep{dam+98},
implying that if it were associated with the bow shock, 
approximately 1.6\% of the energy is used to illuminate
the dust, which is a reasonable fraction to consider (note the total
luminosity from the head of the bow shock 
is $\sim 3.8\times 10^{32}$ erg s$^{-1}$ at the distance of 1.5 kpc). 
On the other hand, the pulsar's kinematic energy can also contribute
to dust heating. Considering a radius $r=0.15$~pc (the head region we have
defined when the distance is 1.5 kpc) interaction area, the energy
received by the ambient medium is 
$(1/2\rho V^2)\cdot V\cdot(\pi r^2)\approx 6.4\times 10^{33}n_H V_{200}^3$ erg~s$^{-1}$, 
where $n_H\simeq 1$ cm$^{-3}$ and $V= 200$ km~s$^{-1}$, the pulsar's velocity, 
are
used. Therefore there is sufficient energy to illuminate the bow shock.

The obvious challenge to this interpretation 
is \psr's location, which is at least 15\arcsec\  
(or 0.11 pc if at the pulsar's distance) away from the determined area 
for the associated object of the bow shock. However, we note that
a pulsar wind can be highly anisotropic and for peculiar cases, 
long jet-like outflows have been detected from pulsars
\citep{jw10,del+11}. Calculations for such anisotropic wind cases 
to produce bow shocks have been considered (e.g., \citealt{wil00}).
One feature that has been revealed is the off-axis location
of an associated star if the star's wind points away from the direction
of the stellar motion, for which the star's location would otherwise be 
determined
to lie in the apparent axis of symmetry when simply fitting the stand-off
region \citep{wil00}. 
\citet{vig+07} did further detailed simulations by considering 
the anisotropy of pulsar winds for pulsar bow shocks, and have shown
that a variety of shape changes with respect to the `standard' shape of
bow shocks.  In particular, the broadening
of the tails of the bow shocks in their simulations SA and SB (see Figure 13 in
their paper), when a pulsar runs through a wall of the high-density ISM 
with finite width, is possibly observed in our case (see Figure~\ref{fig:bow}).
We thus did test fitting to our cometary nebula by using equation A11 given
in \citet{vig+07}, an analytic solution for jet-like outflows
($\lambda$, $c_0$, and $c_2$ were simply set to be 0, 0, and 5, where
$\lambda$ sets the angle between the directions of the pulsar's wind and its 
motion; for details see \citealt{vig+07}). 
The best-fit solution ($\chi^2=13$ for 15
degrees of freedom) is shown in Figure~\ref{fig:bow}, which is nearly
the same as that found in Section~\ref{subsec:loo}. The allowed
3$\sigma$ region for the associated object is however substantially 
elongated toward 
\psr. While the pulsar is still 5.8\arcsec\  away from the region, the
fitting results suggest that it may be possible for the pulsar to drive
the bow shock if it had a highly-collimated wind and the wind
were tilted away from the direction of the pulsar's motion. 
In order to fully
explore this possibility, detailed hydrodynamic simulations, such as
those carried out by \citet{vig+07}, are needed.

In addition, if \psr\ did have a highly collimated pulsar wind, the interaction
of its wind with the ambient medium could be detectable at X-ray energies.
X-ray imaging should be carried out to search for the interaction, which
would lie from \psr\ to the stand-off region of the IR nebula,
thus verifying the association of the pulsar with the nebula. The bow shock
has a direction towards the Galactic plane (Figure~\ref{fig:field}), implying
that the associated object is moving in the same direction. 
High-spatial resolution
radio imaging of the pulsar, to detect its proper motion, should also be
carried out. If \psr\ is also moving in this direction, its chance of
being associated with the bow shock would be substantially increased.

\section{Summary}
\label{sec:sum}

We have discovered an IR bow shock around the pulsar \psr,
and from our follow-up imaging and spectroscopic observations and
WISE all-sky survey data,
a composite spectrum of its head region from 8 to 70 $\mu$m, 
which includes a \spitzer\ IRS 7.5--14.4 $\mu$m spectrum, was obtained. 
In the IRS spectrum, a few PAH and H$_2$ emission features were detected.
Considering the overall spectrum with these features, the emission likely
arises from the dust grains in the interaction region heated by 
collisions and/or radiation flux from the associated object. The dust mass in
the region was estimated to be 2$\times$10$^{29}$~g. We analyzed the shape
of the bow shock by fitting with a standard analytic solution for such objects,
and the derived region in which the associated object would be located is
15\arcsec\ away from \psr. This raises the question of which object
drives this bow shock. No stars were detected in the associated object region 
down to $R\sim 23$,
and moreover our studies of the in-field stars indicate that none of them are 
likely bright O/B stars or red giants located near the Galactic plane, 
which are able to
drive such bow shocks due to their strong winds or mass ejection.
Given the considered properties of pulsar winds, including a few peculiar 
pulsar jet-like outflows, we suggest that it is possible for \psr\ to be
associated with the IR bow shock if it had a highly collimated wind. 
Hydrodynamic simulations of the interaction of such a wind with the ISM
can help explore this possibility. In addition,
X-ray imaging to detect the interaction of its wind with the ambient medium
and high-spatial resolution radio imaging to detect the proper motion
of the pulsar will help verify their association.

\acknowledgements
This work is based in part on observations made with the \textit{Spitzer} 
Space Telescope, which is operated by the Jet Propulsion Laboratory, 
California  Institute of Technology under a contract with NASA.
The publication makes use of data products from the Wide-field Infrared  
Survey Explorer, which is a joint project of the University of California,  
Los Angeles, and the Jet Propulsion Laboratory/California Institute of  
Technology, funded by NASA. 

This research was supported by National Basic Research Program of China
(973 Project 2009CB824800), and National Natural Science Foundation of 
China (11073042). ZW is a Research Fellow of the 
One-Hundred-Talents project of Chinese Academy of Sciences.
PS acknowledges partial support from NASA Contract NAS8-03060.
VMK holds a Canada Research Chair and
the Lorne Trottier Chair in Astrophysics \& Cosmology, and is a
Fellow of the Royal Society of Canada.

{\it Facilities:} \facility{Magellan (PANIC, IMACS), \textit{Spitzer} (IRAC, IRS)} 


\begin{deluxetable}{llccc}
\tablecolumns{5}
\tablecaption{Observations and flux measurements of the cometary nebula around \psr \label{tab:obs}}
\tablewidth{0pt}
\tablehead{
\colhead{Observation date} & \colhead{Telescope/Instrument} & 
\colhead{Band/Wavelength} &  \colhead{Exposure} & Flux measurement\\
\colhead{} & \colhead{} & \colhead{($\mu$m)} & \colhead{(min)} & 
\colhead{(mJy)}}
\startdata
2006 May 17 & Magellan/PANIC & $K_s$ & 22.5 & \nodata \\
2007 Sept 12 & Spitzer/IRAC & 4.5 & 26.8 & $<$0.8 \\
	&		& 8.0 & 26.8 & 10.8$\pm$1.6 \\
2008 June 15 & Magellan/IMACS &	 $R$	& 4.0  & \nodata \\
	&	& H$_{\alpha}$ & 18.3\tablenotemark{a} & \nodata \\
2008 Aug 30 & Spitzer/MIPS & 24 & 11.8 & 20.9$\pm$0.9 \\
	&	& 70 & 10.5 & 82$\pm$32 \\
2009 Sept 14 & Spitzer/IRS & 7.5--14.4 & 32.5 & \nodata \\
2010 Feb 22--25 & WISE & 3.4 & 1.8 & \nodata \\
	&		& 4.6 & 1.8 & \nodata \\
	&		& 12 & 1.8 & 11.4$\pm$0.6 \\
	&		& 22 & 1.8 & 21.1$\pm$1.6 \\
\enddata
\tablenotetext{a}{The surface flux upper limit in H$_\alpha$ band is 
given in \S~\ref{subsec:ha}.}
\end{deluxetable}

\end{document}